\documentclass{article}

\usepackage[]{amsmath,amssymb,amsfonts}
\usepackage[]{tikz}
\usepackage[]{url}
\usepackage{graphicx}

\usepackage{vmargin} 
\setpapersize{A4}
\setmargins{3.5cm}
					 {1.5cm}
           {14.7cm}
           {23.42cm}
           {14pt}
           {1cm}
           {0pt}
           {2cm}

\usepackage{hyperref}
\hypersetup{pdfstartview={XYZ null null 0.85}, pdfstartpage=1}

\usepackage{booktabs}	  	

\usepackage{titlesec}

\usepackage[nottoc,numbib]{tocbibind}

\newtheorem{theorem}{Theorem}

\numberwithin{equation}{section}

\begin{document}

\title{Projection Method for Solving Stokes Flow}

\author{%
  Ryan Hermle
}


\date{\today}
\maketitle
\begin{center}


\begin{abstract}

Various methods for numerically solving Stokes Flow, where a small Reynolds number is assumed to be zero, are investigated.  If pressure, horizontal velocity, and vertical velocity can be decoupled into three different equations, the numerical solution can be obtained with significantly less computation cost than when compared to solving a fully coupled system.  Two existing methods for numerically solving Stokes Flow are explored:  One where the variables can be decoupled and one where they cannot.  The existing decoupling method the limitation that the viscosity must be spatially constant.  A new method is introduced where the variables are decoupled without the viscosity limitation.  This has potential applications in the modeling of red blood cells as vesicles to assist in storage techniques that do not require extreme temperatures, such as those needed for cyropreservation.

\end{abstract}

\vspace{\fill}

rhermle@gmail.com\\
Department of Applied Mathematics, Univ. of Washington

\end{center}


\newpage

\section*{Acknowledgements}
Thank you to all of the excellent professors I have had at UW.  It has been an amazing experience.  A special thank you goes to Chris Vogl*, for believing in me and for giving me the opportunity to do this research with him.

\vspace{\fill}
\begin{center}
*chris.j.vogl@gmail.com\\
Center for Applied Scientific Computing, LLNL
\end{center}
\newpage



\tableofcontents

\newpage

\section{Introduction}

\begin{figure}
	\begin{center}
		\includegraphics[width=13cm]{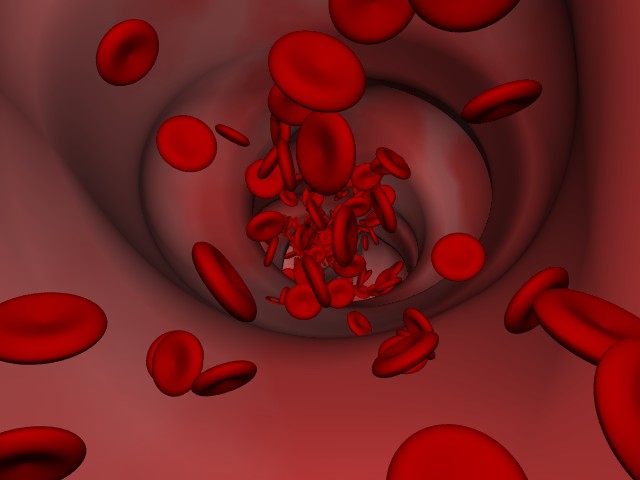}
	\end{center}
	\caption[.]{Red Blood Cells in Fluid \cite{bloodCellImage}}
	\label{artery}
\end{figure}


The fluid dynamics of red blood cells are of great interest to the scientific community.  Biological research \cite{bioResearch} has discovered organisms that can fill their cells with a specific sugar, increasing the viscosity of the intercellular fluids to such high levels that cell metabolism effectively stops, putting the cell in a state of hibernation.  It has been hypothesized that this same method of hibernation could be applied to red blood cells.  This would have a tremendous impact on the ability to store blood for critical medical procedures, such as blood transfusions, as the existing method of cryopreservation requires extremely low temperatures to preserve the blood cells.
\par
\begin{figure}
	\begin{center}
		\includegraphics[width=13cm]{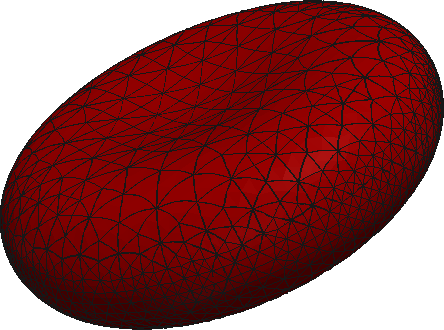}
	\end{center}
	\caption{Rough Geometric Shape of a Red Blood Cell}
	\label{ellipsoid}
\end{figure}

Efficiently modeling the interaction between fluid velocity, pressure, viscosity, and body forces will catalyze the advancement of this groundbreaking storage technique.  In order to begin modeling the fluid dynamics, some preliminary simplifying assumptions are made:

\begin{enumerate}
\item
The cell, being roughly the shape of an ellipsoid as seen in figure \ref{ellipsoid}, exhibits radial symmetry, and thus its dynamics are modeled as a two-dimensional cross section.
\item
The dynamics of the inertial forces of the fluid flow, such as the beating of a heart or the stirring of the fluid, are significantly weaker than those of the viscous forces.
\item
The red blood cells are modeled, for the purposes of this project, as a vesicle, which consists only of the outer membrane of the cell.
\item
The fluid surrounding and inside of the vesicle is incompressible, sharing similar properties to water.

\end{enumerate}

\par
The Reynolds Transport Theorem is used to obtain the equations for conservation of mass and conservation of momentum, which are combined to form the incompressible Navier Stokes Equation: 

\begin{equation}
\label{fullNS}
\rho (\mathbf{u_t} + \nabla \mathbf{u} \cdot \mathbf{u}) = -\nabla p + \nabla \cdot (\mu (\nabla \mathbf{u} + \nabla \mathbf{u}^T)) + \mathbf{f} 
\end{equation}
where $\rho$ represents the density, p represents the pressure, and the vectors $\mathbf{u}$ and $\mathbf{f}$ are defined as:

\begin{equation}
\begin{aligned}
\mathbf{u} = \begin{bmatrix} u \\ v \end{bmatrix} \\
\mathbf{f} = \begin{bmatrix} f1 \\ f2 \end{bmatrix}. \\
\end{aligned}
\end{equation} 
Here, $u$ and $v$ represent the horizontal velocity of the fluid, $f1$ and $f2$ represent the horizontal and vertical components of the force.  $-\nabla p$ represents the force from the pressure on the fluid, $\nabla \cdot (\mu (\nabla \mathbf{u} + \nabla \mathbf{u}^T))$ represents the viscous force, $\rho (\mathbf{u_t} + \nabla \mathbf{u} \cdot \mathbf{u})$ represents the inertial force, and $\mathbf{f}$ represents external body forces, such as gravity or the force from a cell wall.  
\par
The incompressibility is imposed through the conservation of mass equation:

\begin{equation}
\frac{\partial \rho}{\partial t} + \nabla \cdot (\rho \mathbf{u}) = 0.
\end{equation}
Because the density cannot change, $\frac{\partial \rho}{\partial t} = 0$.  It is assumed that $\rho$ is also constant with respect to space, so $\nabla \cdot (\rho \mathbf{u}) = \nabla \rho \cdot \mathbf{u} + \rho \nabla \cdot \mathbf{u} = 0 + \rho \nabla \cdot \mathbf{u}$.  This yields the divergence-free condition:

\begin{equation}
\label{divZero}
\nabla \cdot \mathbf{u} = 0
\end{equation}
which will be a very important result throughout this paper.
\par
To properly model the behavior of interest, the full Navier Stokes Equation \ref{fullNS} is non-dimensionalized to reflect the idea that the viscous forces are much greater than the inertial forces.  Variables with a tilde over them indicate characteristic quantities:

\begin{equation}
\begin{aligned}
&\mathbf{x}^* = \frac{\mathbf{x}}{\tilde{L}} \hspace{1cm} (\Rightarrow \nabla = \tilde{L}^{-1} \nabla^*) \\
&t^* = \frac{t}{\tilde{T}}, \text{ with } \tilde{T} = \frac{\tilde{L}}{\tilde{u}} \hspace{1cm}  (\Rightarrow \frac{\partial}{\partial t} = \tilde{T}^{-1} \frac{\partial}{\partial t^*})\\
&\mathbf{u}^* = \frac{\mathbf{u}}{\tilde{u}}\\
&p^* = \frac{p}{\tilde{p}}, \text{ with } \tilde{p} = \frac{\mu}{\tilde{T}}\\
&\mathbf{f} = \frac{\mathbf{f}}{\tilde{f}}, \text{ with } \tilde{f} = \frac{\mu}{\tilde{L}\tilde{T}}.
\end{aligned}
\end{equation}
Substituting these terms into equation \ref{fullNS} and simplifying yields

\begin{equation}
\label{rawNonD}
\rho \tilde{L} \tilde{u} \mu^{-1} (\mathbf{u}_{t^*}^* + \nabla^* \mathbf{u}^* \cdot \mathbf{u}^*) = - \nabla^* p^* + \nabla^* \cdot (\mu(\nabla^* \mathbf{u}^* + \nabla^* \mathbf{u}^{*T})) + \mathbf{f}^*.
\end{equation}
Defining the Reynolds Number as

\begin{equation}
\label{Re}
Re = \frac{\rho \tilde{u}}{\mu / \tilde{L}}
\end{equation}
and dropping the star notation converts equation \ref{rawNonD} to

\begin{equation}
Re(\mathbf{u}_t + \nabla \mathbf{u} \cdot \mathbf{u}) = - \nabla p + \nabla \cdot (\mu (\nabla \mathbf{u} + \nabla \mathbf{u}^T)) + \mathbf{f}.
\end{equation}
\par
Physically speaking, the Reynolds number from equation \ref{Re} represents the ratio of the inertial forces to viscous forces.  Thus, in line with the prior assumptions, the dominant viscous forces result in a small Reynolds Number.  As the Reynolds Number approaches zero, the full Navier Stokes equation from \ref{fullNS} reduces to:

\begin{equation}
\label{reducedNS}
0 = -\nabla p + \nabla \cdot (\mu ( \nabla \mathbf{u} + \nabla \mathbf{u}^{T})) + \mathbf{f}.
\end{equation}
\par
Define the Laplace Operator as

\begin{equation}
\Delta g = g_{i,jj}
\end{equation}
If the viscosity is spatially constant ($\mu(\mathbf{x}) = \mu$), the $\nabla \cdot (\mu (\nabla \mathbf{u} + \nabla \mathbf{u}^T))$ term becomes just $\mu \Delta \mathbf{u}$. The $\nabla \cdot \nabla \mathbf{u}^T$ term goes to zero because of the divergence-free condition:

\begin{equation}
\begin{aligned}
\nabla \cdot (\nabla \mathbf{u})^T &= \nabla \cdot (u_{i,j})^T = \nabla \cdot \begin{bmatrix} u_x & u_y \\ v_x & v_y \end{bmatrix}^T\\ 
&= \nabla \cdot \begin{bmatrix} u_x & v_x \\ u_y & v_y \end{bmatrix} = \begin{bmatrix} u_{xx} + v_{xy} \\ u_{yx} + v_{yy} \end{bmatrix}\\
&= \begin{bmatrix} (u_x + v_y)_x \\ (u_x + v_y)_y \end{bmatrix} = \begin{bmatrix} 0 \\ 0 \end{bmatrix}\\
\end{aligned}
\end{equation}
Thus, when $\mu$ is spatially constant, equation \ref{reducedNS} becomes:

\begin{equation}
\label{reducedNS2}
0 = -\nabla p + \mu \Delta \mathbf{u} + \mathbf{f}.
\end{equation}

\section{Existing Methods}

\subsection{Saddle-Point Method}  

For the Saddle-Point Method, the velocity and pressure are simultaneously solved as the following system of equations:

\begin{equation}
\begin{aligned}
-\nabla p + \mu \Delta \mathbf{u} &= -\mathbf{f}\\
\nabla \cdot \mathbf{u} &= 0
\end{aligned}
\end{equation}

\begin{figure}
	\begin{center}
		\includegraphics[width=6cm]{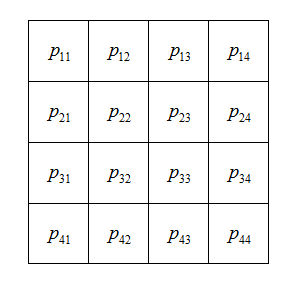}
	\end{center}
	\caption{Pressure Grid, $M=4$}
	\label{pGrid}
\end{figure}
\par
To perform this task, $p$, $u$, and $v$ are discretized and then stacked into a single vector, denoted as $\mathbf{x}$.  The number of discretization points in a single direction, defined as M, will determine the number of rows in $\mathbf{x}$.  Since the domain is two-dimensional, each field is represented by $M \times M$ discretized values to span the entire domain.  The grid for pressure is visualized in Figure \ref{pGrid}.  The grids for $u$ and $v$ take on the same form.  Thus, $\mathbf{x}$ has a total of $3M^2$ rows, with $M^2$ rows for each of $p$, $u$, and $v$:

\begin{equation}
\mathbf{x} = \begin{bmatrix} p_{11} \\ ... \\p_{M1} \\ ...\\ p_{MM} \\ u_{11} \\ ... \\ u_{MM} \\ v_{11} \\ ... \\ v_{MM} \end{bmatrix},
\end{equation} 
where $p_{i+1,j}$ is $\Delta x$ to the right of $p_{i,j}$, and $p_{i,j+1}$ is $\Delta y$ above $p_{i,j}$.
\par
Next, a matrix $A$ is constructed to approximate the differential operators acting on each discretized variable of $p$, $u$, and $v$ using the second-order finite-difference formulas: 
\par
$O(\Delta t^2)$ Center-Difference Schemes:
\begin{equation}
f^{'}(t) \approx [f(t + \Delta t) - f(t - \Delta t)] / 2 \Delta t
\end{equation}
\begin{equation}
f^{''}(t) \approx [f(t + \Delta t) - 2f(t) + f(t - \Delta t)] / \Delta t^2
\end{equation}
\par
$O(\Delta t^2)$ Forward- and Backward-Difference Schemes:
\begin{equation}
f^{'}(t) \approx [-3f(t) + 4f(t + \Delta t) - f(t + 2 \Delta t)] / 2 \Delta t
\end{equation}
\begin{equation}
f^{'}(t) \approx [3f(t) - 4f(t - \Delta t) + f(t - 2 \Delta t)] / 2 \Delta t
\end{equation}
The size of $A$ is $3M^2 \times 3M^2$.  Each row represents one equation for one of the discretized variables.  This creates a large system 

\begin{equation}
\mathbf{A} \mathbf{x} = \mathbf{b}
\end{equation}
which has the form

\begin{equation}
\label{saddleMatrix}
\begin{bmatrix} - G & \mu L \\ 0 & D \end{bmatrix} \begin{bmatrix} p \\ \mathbf{u} \end{bmatrix} = \begin{bmatrix} -\mathbf{f} \\ 0 \end{bmatrix}.
\end{equation}
where G,L, and D are discretized versions of the gradient, Laplacian, and divergence operators.  In order to properly index through $3M^2$ rows and columns, the intuitive two-dimensional subindices can be mapped to a single index using the MATLAB function \textbf{sub2ind}, and then $M^2$ points can be added to the linear index to get to the next variable.  A more robust way to map the indices, which is used in most of the algorithms presented, is to create an $M \times M$ mapping matrix for each variable that stores the single index corresponding to each two-dimensional subindex for that variable.  
\par
The matrices $\mathbf{A}$ and $\mathbf{b}$ are constructed by looping through each of the $3M^2$ rows and applying the correct finite differences that apply to the discretized variables that are represented in that row.  Once the matrices are constructed, the system is easily solved using the MATLAB \verb|"\"| operator, also known as \textbf{mldivide}.  From there, the solved values of $p$, $u$, and $v$ are extracted.

\subsection{Fluid in a Pipe}
\label{fluidInPipeSection}

The Saddle-Point Method, as well as methods presented later, is used to model fluid flow in a pipe.  The flow is driven by a pressure difference between the left and right edges of the pipe. It is assumed that there are no body forces, so the only force will be in the horizontal direction from the pressure difference.  The vertical velocity $v$ is assumed to be zero.
\par
\begin{figure}
	\begin{center}
		\includegraphics[width=13cm]{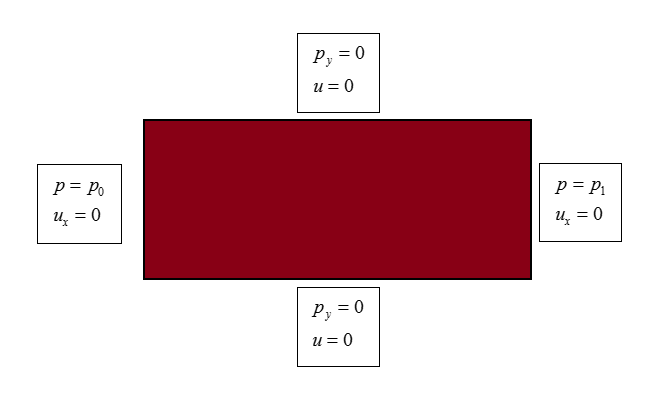}
	\end{center}
	\caption{Boundary Conditions for Fluid in a Pipe}
	\label{BC1}
\end{figure}

A no-slip top and bottom boundary condition is applied, assuming that there is adhesion between the wall of the pipe and the fluid such that the horizontal velocity of the fluid is zero at the top and bottom of the pipe.  The pressure is expected to change in the horizontal direction due to the applied gradient, but it is not expected to change in the vertical direction, so a Neumann boundary condition of $p_y = 0$ is applied at the top and bottom of the pipe.  Finally, $u$ is expected to change in the vertical direction due to the no-slip condition on the walls of the pipe, but it is not expected to vary horizontally, so a Neumann boundary condition of $u_x = 0$ is applied to the left and right edges of the domain.  A summary of these conditions is depicted in Figure \ref{BC1}.
The values for $p_0$, $p_1$, and $\mu$ are chosen arbitrarily to be $200$, $100$, and $2$, respectively.  These boundary conditions are then placed into the matrix $\mathbf{A}$ from equation \ref{saddleMatrix}, using the indices of the discretized variables that reside near the boundaries.  One-sided difference formulas are used for the Neumann conditions.

\subsection{Checkerboarding and the Staggered Grid} 

\begin{figure}
	\begin{center}
		\includegraphics[width=13cm]{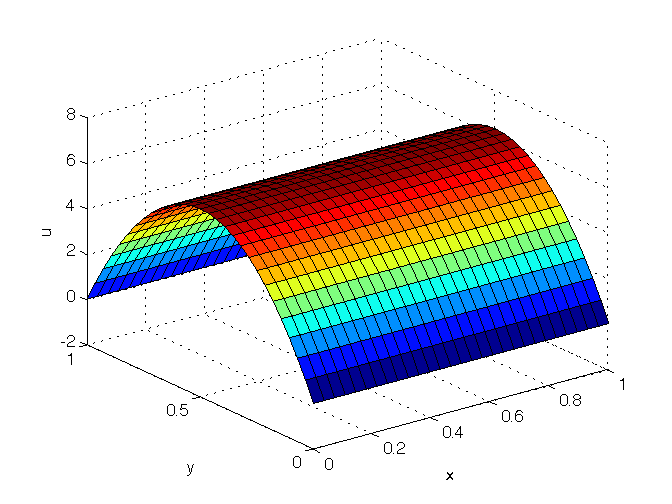}
	\end{center}
	\caption{Surface Plot of Horizontal Velocity}
	\label{uChecker}
\end{figure}

\begin{figure}
	\begin{center}
		\includegraphics[width=13cm]{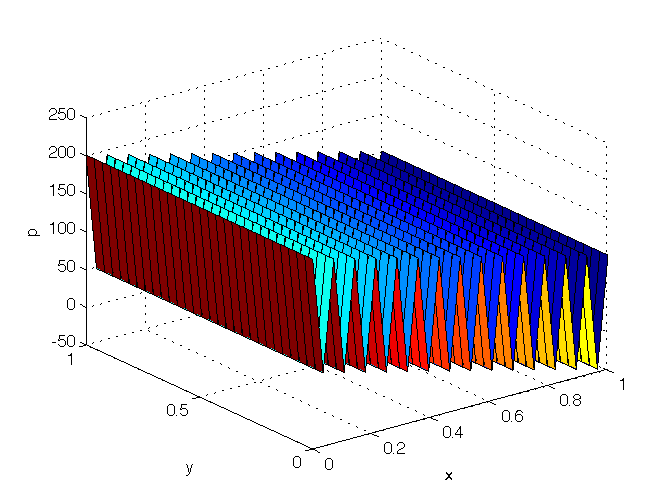}
	\end{center}
	\caption{Surface Plot of Pressure, Showing the Checkerboard Pattern}
	\label{pChecker}
\end{figure}

Applying the Saddle-Point method to the setup specified in section \ref{fluidInPipeSection} produces the surface plots for $p$ and $u$ shown in Figures \ref{uChecker} and \ref{pChecker}.  The surface plot for $u$ looks as expected:  zero near the walls of the pipe due to the no-slip condition and a parabolic profile caused by the pressure gradient.  However, the pressure graph depicts a clear problem.  It seems to be very inconsistent, not smooth, and oscillating wildly.  This is what is known as a checkerboard pattern, and it arises from the way the pressure has been discretized and the particular finite difference formulas used.
\par
To explain this phenomenon, consider the following half of equation \ref{saddleMatrix} written in component form and with zero body force:

\begin{equation}
\mu (u_{xx} + u_{yy}) = p_x
\end{equation}
and recall that the second-order finite difference formula used to represent $p_x$ is

\begin{equation}
p_x \approx [p(x + \Delta x) - p(x - \Delta x)] / 2 \Delta x.
\end{equation}
The key detail is that the pressure on the left boundary $p_0$, which helps determine the pressure gradient for the entire system, is only connected to every other pressure node due to how the finite difference formula skips over the node that is being approximated.  This creates a chain connecting every other pressure node horizontally across the entire domain.  This can be seen in Figure \ref{pChecker}, as every other pressure node looks correct, smoothly transitioning from the left boundary condition to the right boundary condition, while the other nodes take on a different value because they are not connected to the boundary conditions.
\par
\begin{figure}
	\begin{center}
		\includegraphics[width=10cm]{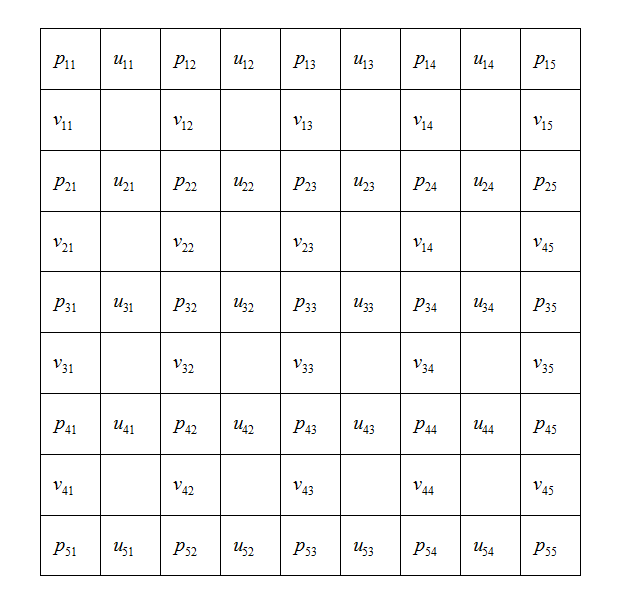}
	\end{center}
	\caption{Staggered Grid for p, u, and v}
	\label{saddleStaggered}
\end{figure}

In order to address this issue, a staggered grid is created as illustrated in Figure \ref{saddleStaggered}.
This orientation of cells allows the use of a modified second-order finite-difference formula that uses consecutive grid values of $p$ to calculate the first derivative. To derive this formula, Taylor expansions are done about the $u$ and $v$ nodes, using the p cells just to the sides of these nodes with only a distance of $\frac{\Delta x}{2}$.  For the sake of simplicity, it is assumed that the grid spacing is created such that $\Delta x = \Delta y$:

\begin{equation*}
\begin{aligned}
p(x + \frac{\Delta x}{2},y) = p(x,y) + \frac{\Delta x}{2} \frac{\partial p(x,y)}{\partial x} + O(\Delta x^2)\\
p(x - \frac{\Delta x}{2},y) = p(x,y) - \frac{\Delta x}{2} \frac{\partial p(x,y)}{\partial x} + O(\Delta x^2)\\
\end{aligned}
\end{equation*}
Subtracting these two equations and simplifying yields

\begin{equation*}
p_x = \frac{[p(x + \frac{\Delta x}{2},y) - p(x - \frac{\Delta x}{2},y)]}{ \Delta x} + O(\Delta x^2).
\end{equation*}
The same logic is applied for $p_y$:

\begin{equation*}
p_y = \frac{p(x, y + \frac{\Delta y}{2}) - p(x, y - \frac{\Delta y}{2},y)}{ \Delta y} + O(\Delta y^2).
\end{equation*}
In index form, these finite difference formulas can be written as follows:

\begin{equation*}
p_x \approx \frac{p_{i+1,j} - p_{i,j}}{ \Delta x}
\end{equation*}

\begin{equation*}
p_y \approx \frac{p_{i,j+1} - p_{i,j}}{ \Delta y}
\end{equation*}
The same logic is applied to calculate derivatives of $\mathbf{u}$ at a $p$ cell, but the indices are shifted by one because of the way the grid is structured:

\begin{equation*}
u_x \approx \frac{u_{i,j} - u_{i-1,j}}{ \Delta x}.
\end{equation*}

\begin{equation*}
v_y \approx \frac{v_{i,j} - v_{i,j-1}}{ \Delta y}.
\end{equation*}

\begin{figure}
	\begin{center}
		\includegraphics[width=13cm]{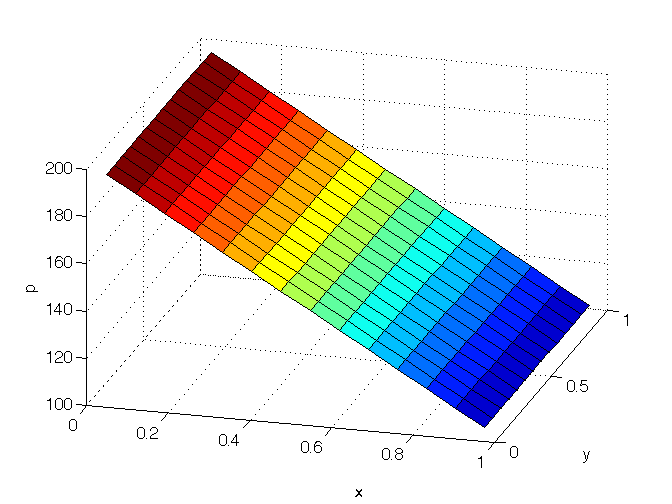}
	\end{center}
	\caption{Correct Pressure Field Using the Staggered Grid}
	\label{pStaggered}
\end{figure}

\begin{figure}
	\begin{center}
		\includegraphics[width=13cm]{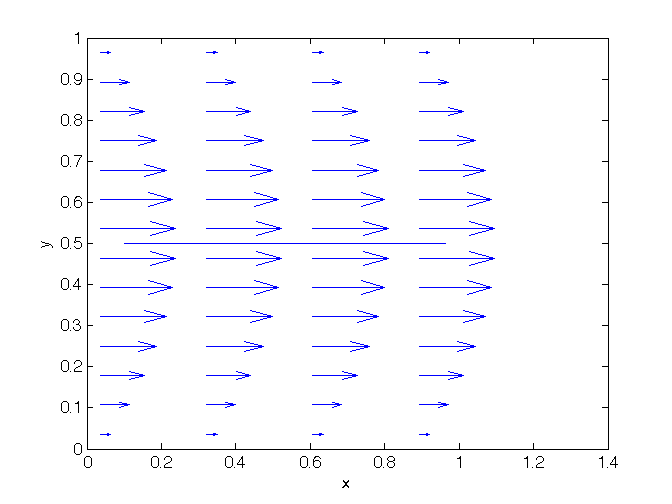}
	\end{center}
	\caption{Vector Field for Fluid in a Pipe}
	\label{saddlePipeQuiver}
\end{figure}

This algorithm will connect each cell in a contiguous chain, so the first derivative formula will no longer skip every other cell.  Other operators that do not involve cells of a different type are unchanged, using the normal finite-difference formulas. This modification produces the correct pressure surface plot, as shown in Figure \ref{pStaggered}.  A quiver plot, which shows the magnitude and direction of the velocity vector $\mathbf{u}$ on a two-dimensional plane, is also shown in Figure \ref{saddlePipeQuiver}.

\subsection{Decoupling Method}  

While the Saddle-Point method correctly solved the fluid in a pipe problem, it is computationally expensive.  An alternate method utilizes a divide and conquer approach to increase efficiency.
\par
Taking the divergence of both sides of the Stokes Equation (\ref{reducedNS}) yields interesting results:  

\begin{equation}
\begin{aligned}
0 &= \nabla \cdot (- \nabla p + \mu \Delta \mathbf{u} + \mathbf{f})\\
\implies 0 &= \nabla \cdot \left( - \begin{bmatrix} p_x \\ p_y \end{bmatrix} + \mu \begin{bmatrix} u_{xx} + u_{yy} \\ v_{xx} + v_{yy} \end{bmatrix} + \begin{bmatrix} f1 \\ f2 \end{bmatrix} \right)\\
\implies 0 &= -(p_{xx} + p_{yy}) + \mu (u_{xxx} + u_{yyx} + v_{xxy} + v_{yyy}) + f1_x + f2_y.
\end{aligned}
\end{equation}
By rearranging the mixed partial derivatives and applying the divergence free condition $\nabla \cdot \mathbf{u} = u_x + v_y = 0$, one obtains

\begin{equation}
\begin{aligned}
0 &= -(p_{xx} + p_{yy} + \mu ((u_x + v_y)_{xx} + (u_x + v_y)_{yy}) + f1_x + f2_y \\
\implies 0 &= -(p_{xx} + p_{yy}) + f1_x + f2_y \\
\implies \Delta p &= \nabla \cdot \mathbf{f}.
\end{aligned}
\end{equation} 
As a result, the pressure and velocity are decoupled into the following three equations:

\begin{equation}
\label{decoupledSystem1}
p_{xx} + p_{yy} = f1_x + f2_y
\end{equation}
\begin{equation}
\label{decoupledSystem2}
\mu (u_{xx} + u_{yy})  = p_x - f1
\end{equation}
\begin{equation}
\label{decoupledSystem3}
\mu (v_{xx} + v_{yy})  = p_y - f2
\end{equation}
This very effective and simple decoupling of the pressure and velocity depends on the fact that no extra terms are generated from applying derivatives to the spatially constant viscosity term $\mu$.  Note, to fully model the behavior of the vesicles in the scenario of interest, the viscosity is not spatially constant, so this method cannot be used in that case.  
\par
\begin{figure}
	\begin{center}
		\includegraphics[width=13cm]{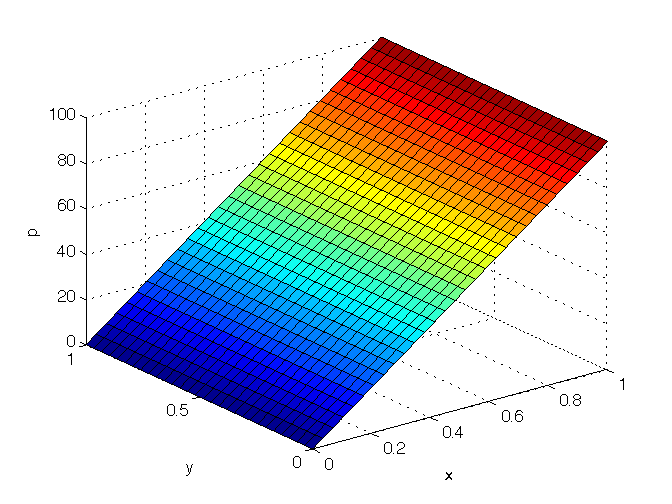}
	\end{center}
	\caption{Pressure Field Generated from the Decoupling Method}
	\label{pShortcut}
\end{figure}
\begin{figure}
	\begin{center}
		\includegraphics[width=13cm]{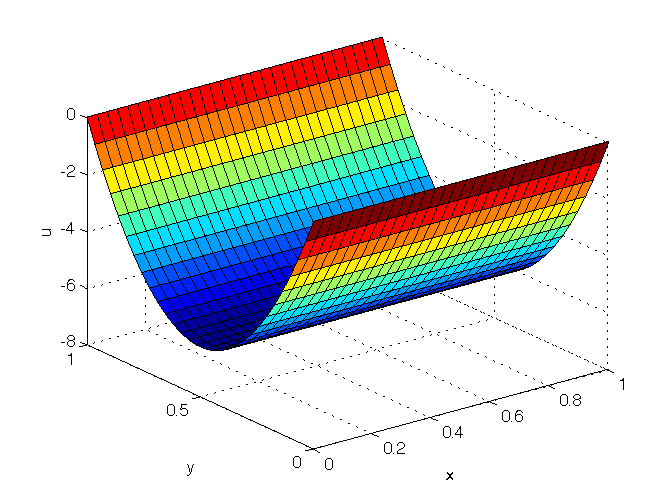}
	\end{center}
	\caption{Horizontal Velocity Field Generated from the Decoupling Method}
	\label{uShortcut}
\end{figure}

The solution is obtained using three linear system solves of the form $\mathbf{A} \mathbf{x} = \mathbf{b}$, where each vector $\mathbf{x}$ now only contains $M^2$ rows, and $\mathbf{A}$ is now $M^2 \times M^2$.  The matrix $\mathbf{A}$ in each of the solves is still constructed by looping through the $M^2$ rows and assigning the finite differences that apply.  Solving the system created by the boundary conditions from the fluid in a pipe scenario yields the smooth pressure and velocity graphs as seen in Figures \ref{pShortcut} and \ref{uShortcut}.  In this case, the values for $p_0$, $p_1$, and $\mu$ are chosen as $0$, $100$, and $2$, respectively.

\subsection{Analytic Solution for Fluid in a Pipe}  

In order to verify that the algorithms are working correctly, the analytic solution for this problem is easily produced.  The calculations are especially easy if the domain is defined as $x \in [0,1]$ and $y \in [0,1]$.  The decoupled system of equations \ref{decoupledSystem1} through \ref{decoupledSystem3} are used to derive the solution.
\par
Since the pressure gradient is horizontal and there are no body forces acting on the fluid, an Ansatz can be made that that $p(x,y) = p(x)$.  Thus, $\Delta p = 0$ becomes $p_{xx} = 0$, which is easily solved by integrating twice:

\begin{equation}
\begin{aligned}
&p_{xx} = 0\\
\implies &\int_0^1{p_{xx}} dx = C_1\\
\implies &\int_0^1{p_x} dx = C_2x + C_1\\
\implies &p = C_2x + C_1
\end{aligned}
\end{equation}
Enforcing the boundary conditions that $p = p_1$ on the right side and $p = p_0$ on the left side yields

\begin{equation}
p = p_0 + x(p_1 - p_0).
\end{equation}
A similar Ansatz is made for $u$, the horizontal component of velocity.  Since $u$ varies only with y, the Ansatz is $u(x,y) = u(y)$.  Noting that the $p_x$ component of $\nabla p$ can be easily calculated using the analytic solution.  A simplified equation for $u$ is found:

\begin{equation}
\begin{aligned}
&\mu u_{yy} = p_x\\
\implies &\mu u_{yy} = p_1 - p_0
\end{aligned}
\end{equation}
This equation can also be easily solved by integrating twice:

\begin{equation}
\begin{aligned}
&\mu \int_0^1{u_{yy}} dy = (p_1 - p_0)y + C_1\\
\implies &\mu \int_0^1{u_y} dy = \frac{(p_1 - p_0)}{2} y^2 + C_1y + C_2
\end{aligned}
\end{equation}
Enforcing the no-slip boundary conditions, specifically that $u=0$ at the top and bottom boundaries, yields the analytic solution for $u$:

\begin{equation}
u = \frac{1}{2 \mu} (p_1-p_0)y(y-1)
\end{equation}
Additionally, it has already been noted that $v=0$.

\subsection{Convergence Testing}

\begin{figure}
	\begin{center}
		\includegraphics[width=13cm]{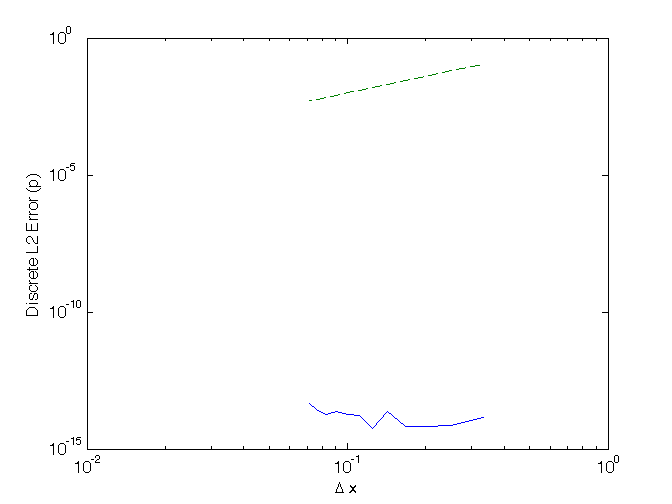}
	\end{center}
	\caption{Discrete L2 Error for Pressure.  The dashed line is a reference line for second-order.}
	\label{L2EPSaddlePipe}
\end{figure}

\begin{figure}
	\begin{center}
		\includegraphics[width=13cm]{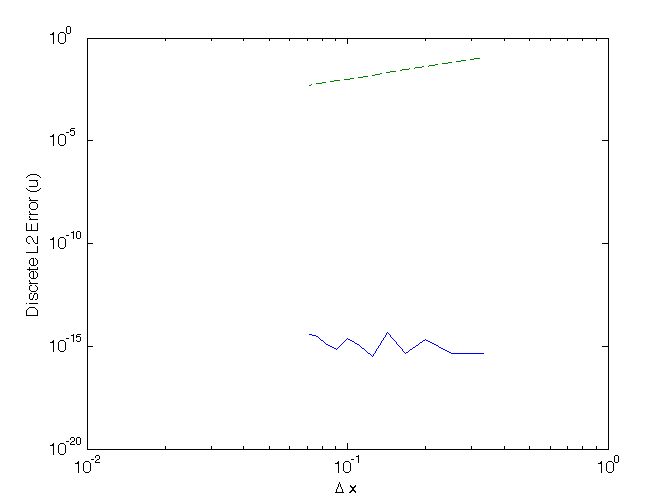}
	\end{center}
	\caption{Discrete L2 Error for Horizontal Velocity}
	\label{L2EUSaddlePipe}
\end{figure}

\begin{figure}
	\begin{center}
		\includegraphics[width=13cm]{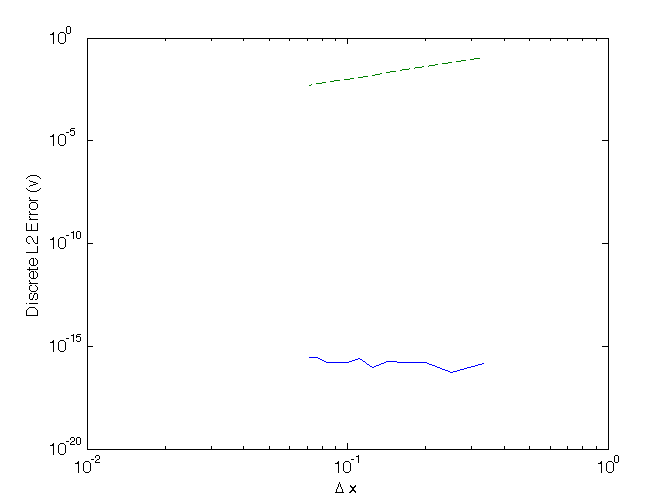}
	\end{center}
	\caption{Discrete L2 Error for Vertical Velocity}
	\label{L2EVSaddlePipe}
\end{figure}

The numerically solved values for $u$, $v$, and $p$ are compared against the values of the analytic solution using the discrete L2 error.  Since second-order finite-difference formulas are used in each algorithm, the discrete L2 error should be $O(\Delta x^2)$.  The error calculation is defined in the following equations

\begin{equation}
E_{p} = \sqrt{\frac{\sum_{i=1}^M \sum_{j=1}^M (p_{ij} - p(x_{ij},y_{ij}))^2}{M^2}}
\end{equation}
\begin{equation}
E_{u} = \sqrt{\frac{\sum_{i=1}^M \sum_{j=1}^M (u_{ij} - u(x_{ij},y_{ij}))^2}{M^2}}
\end{equation}
\begin{equation}
E_{v} = \sqrt{\frac{\sum_{i=1}^M \sum_{j=1}^M (v_{ij} - v(x_{ij},y_{ij}))^2}{M^2}}
\end{equation}
where $p(x_{ij},y_{ij})$, $u(x_{ij},y_{ij})$, and $v(x_{ij},y_{ij})$ represent the values of the analytical solutions calculated at the corresponding spatial locations.
\par
Figures \ref{L2EPSaddlePipe} through \ref{L2EVSaddlePipe} show the discrete L2 error for $p$, $u$, and $v$, calculated using the Saddle-Point method with the staggered grid and values of $200$, $100$, and $2$ for $p_0$, $p_1$, and $\mu$.  Since the analytic solutions for the fluid in a pipe model are a first order polynomial for $p$, a second order polynomial for $u$, and identically 0 for $v$, the second-order finite difference scheme approximates the system with machine-order precision.  The graphs used to present the L2 error are log scaled in both the x-axis and the y-axis, where the x-axis represents the current discretization distance $\Delta x$ and the y-axis represents the discrete L2 error for the respective variable.  
\par
The reference line on the graphs is a line with a slope of 2, created by graphing $\Delta x$ vs $\Delta x^2$ on the log-log scale.  The reasoning behind this is that if the discrete L2 error scales as $\Delta x^2$, graphing $\log(\Delta x)$ on the horizontal axis and $\log(O(\Delta x^2))$ on the vertical axis should result in a line with slope 2.  The reference line is then used to verify the algorithm is converging with second-order precision, which was of course not necessary in this particular example due to the machine-level precision.  Nearly identical values are obtained when using the Decoupling Method for the same scheme.

\section{Projection Method for Stokes Flow}

\subsection{Derivation}  

The existing methods are established and convergence has been shown, but the problem remains that the Saddle-Point method is relatively slow and the Decoupling method cannot solve for spatially varying viscosity.  A method that decouples the pressure solve from the velocity solve but does not assume spatially constant viscosity is needed.  This problem motivates the use of the Projection Method, which begins with what is known as the Helmholtz-Hodge Decomposition Theorem:

\begin{theorem}
A vector field $\mathbf{\Psi}$ defined on a simply connected domain can be uniquely decomposed into a divergence-free component, $\mathbf{\Gamma}$, and a curl-free component, $\nabla{\Phi}$:
\begin{equation}
\label{HD}
\mathbf{\Psi} = \mathbf{\Gamma} + \nabla{\Phi}
\end{equation}
\end{theorem}
The full Navier Stokes Equation (\ref{fullNS}) is rearranged to

\begin{equation}
\label{NSDivCurl}
\nabla \cdot (\mu (\nabla \mathbf{u} + \nabla \mathbf{u}^T)) - Re(\nabla \mathbf{u} \cdot \mathbf{u}) + \mathbf{f} = Re(\mathbf{u}_t) + \nabla p.
\end{equation}
Taking a derivative with respect to time of the divergence-free condition gives

\begin{equation}
\label{utDivFree}
\nabla \cdot \mathbf{u_t} = \frac{\partial}{\partial t} \nabla \cdot \mathbf{u} = \frac{\partial}{\partial t} 0 = 0.
\end{equation}
Thus, equation \ref{NSDivCurl} is in the form of the Helmholtz-Hodge Decomposition (\ref{HD}).  $Re(\mathbf{u}_t)$ is divergence-free, as previously stated, so it is the divergence-free component $\mathbf{\Gamma}$.  Letting $\Phi = p$, $\nabla p$ is the curl-free space.  The left-hand side of the equation represents the vector field $\mathbf{\Psi}$.
\par
Define an inner product between two vector fields $\mathbf{a}$ and $\mathbf{b}$ as
\begin{equation}
<\mathbf{a}, \mathbf{b}> =  \iint\limits_\Omega \mathbf{a} \cdot \mathbf{b} \text{ } dx dy,
\end{equation}
where $\Omega$ is the two-dimensional domain.  The projection of vector field $\mathbf{a}$ onto the curl-free $\nabla p$ is thus defined as

\begin{equation}
proj_{\nabla p} \mathbf{a} = \frac{<\mathbf{a}, \nabla p>}{<\nabla p, \nabla p>} \nabla p.
\end{equation}
Define 

\begin{equation}
\label{P}
P(\mathbf{a}) = \mathbf{a} - proj_{\nabla p} \mathbf{a}.
\end{equation}  
Using this operator, consider $P(\mathbf{u}_t) = \mathbf{u}_t - proj_{\nabla p} \mathbf{u}_t$.  Applying a two-dimensional integration by parts, $<\mathbf{u}_t, \nabla p>$ is given as




\begin{equation}
\iint\limits_\Omega \mathbf{u}_t \cdot \nabla p \text{ } dxdy = \oint\limits_{\partial \Omega} p \mathbf{u}_t \cdot \hat{n} \text{ } dS - \iint\limits_\Omega p \nabla \cdot \mathbf{u}_t \text{ } dxdy.
\end{equation}
Since $\nabla \cdot \mathbf{u}_t = 0$ (equation \ref{utDivFree}), the double integral on the right-hand side vanishes.  If boundary conditions are applied such that $\mathbf{u} \cdot \hat{\mathbf{n}} = 0$ along the boundary, then

\begin{equation}
\oint\limits_{\partial \Omega} p \mathbf{u}_t \cdot \hat{n} \text{ } dS = 0 \implies <\mathbf{u}_t, \nabla p> = 0.
\end{equation}
Thus,

\begin{equation}
\label{Put}
P(\mathbf{u}_t) = \mathbf{u}_t.
\end{equation}
Similar reasoning can be used to show that $P(\mathbf{u}) = \mathbf{u}$ as well.  Note that $P(\nabla p) = 0$:

\begin{equation}
\label{Pp}
\begin{aligned}
P(\nabla p) &= \nabla p - \frac{<\nabla p, \nabla p>}{<\nabla p, \nabla p>} \nabla p\\
						&= \nabla p - (1) \nabla p\\
						&= 0.
\end{aligned}
\end{equation}
\par
\begin{figure}
	\begin{center}
		\includegraphics[width=13cm]{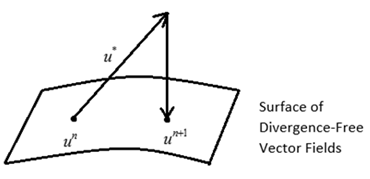}
	\end{center}
	\caption{Illustration of the Projection Method}
	\label{projection}
\end{figure}

These concepts are well known through the work of Alexandre Chorin \cite{chorin}.  The algorithm, originally known as Chorin's projection method, is used to numerically solve the incompressible Navier-Stokes Equation (\ref{fullNS}).  It takes advantage of the fact that the equation can be written in the form of the Helmholtz-Hodge Decomposition.  Its general steps are outlined as follows:
\begin{enumerate}
\item Compute an intermediate velocity, $\mathbf{u^*}$, that ignores the pressure gradient term.  This takes the solution off of the true divergence-free solution space.
\item Use the intermediate velocity to solve for the pressure.
\item Use the pressure and the intermediate velocity to project the solution back onto the divergence-free solution space, obtaining a solution for $\mathbf{u}$ that represents its steady state after $\Delta t$ units of time have passed.
\end{enumerate}
This idea is illustrated in Figure \ref{projection}.


\par
To solve the scenario of interest in this paper, Chorin's projection method is modified in order to obtain a solution to the Stokes Flow equation.  A number of steps must be taken to obtain $\mathbf{u}^{n+1}$ from $\mathbf{u}^{n}$.  Starting with a rearranged version of the full Navier Stokes Equation, the $P$ operator (\ref{P}) is applied to both sides:

\begin{equation}
\label{fullNSRearranged}
Re(\mathbf{u}_t) = -Re(\nabla \mathbf{u} \cdot \mathbf{u}) - \nabla p + \nabla \cdot (\mu (\nabla \mathbf{u} + \nabla \mathbf{u}^T)) + \mathbf{f}
\end{equation}

\begin{equation}
\label{projBothSides}
P(Re(\mathbf{u}_t)) = P(-Re(\nabla \mathbf{u} \cdot \mathbf{u}) - \nabla p + \nabla \cdot (\mu (\nabla \mathbf{u} + \nabla \mathbf{u}^T)) + \mathbf{f})
\end{equation}
Applying the identities $P(\mathbf{u}_t) = \mathbf{u}_t$ (\ref{Put}) and $P(\nabla p) = 0$ (\ref{Pp}) reduces equation \ref{projBothSides} to

\begin{equation}
Re(\mathbf{u}_t) = P(-Re(\nabla \mathbf{u} \cdot \mathbf{u}) + \nabla \cdot (\mu (\nabla \mathbf{u} + \nabla \mathbf{u}^T)) + \mathbf{f}).
\end{equation}
Since $P(\mathbf{u}) = \mathbf{u}$, $\frac{1}{\Delta t} \mathbf{u}$ can be added within the projection operator and then subtracted outside.  $\Delta t$ defines the amount of time between solutions of $\mathbf{u}(\mathbf{x},t)$:

\begin{equation}
\label{addInside}
Re(\mathbf{u}_t) = P\left(-Re(\nabla \mathbf{u} \cdot \mathbf{u}) + \nabla \cdot (\mu (\nabla \mathbf{u} + \nabla \mathbf{u}^T)) + \mathbf{f} + \frac{1}{\Delta t}\mathbf{u}\right) - \frac{1}{\Delta t}\mathbf{u}.
\end{equation}

\begin{equation}
\Delta t = t_{n+1} - t_n
\end{equation}
Both sides of the equation are then integrated in time from $t = t_n$ to $t = t_{n+1}$:

\begin{equation}
\label{integralApprox}
Re(\mathbf{u}^{n+1} - \mathbf{u}^n) = \int_{t_n}^{t_{n+1}}P\left(-Re(\nabla \mathbf{u} \cdot \mathbf{u}) + \nabla \cdot (\mu (\nabla \mathbf{u} + \nabla \mathbf{u}^T)) + \mathbf{f} + \frac{1}{\Delta t}\mathbf{u}\right)dt - \frac{1}{\Delta t}\int_{t_n}^{t_{n+1}}\mathbf{u}dt.
\end{equation}
The first integral is evaluated using a left-hand rectangular approximation, and the second integral is evaluated using a right-hand approximation.  This creates a relationship between $\mathbf{u^n}$ and $\mathbf{u^{n+1}}$:

\begin{equation}
Re(\mathbf{u}^{n+1} - \mathbf{u}^n) = P\left(-Re(\nabla \mathbf{u}^n \cdot \mathbf{u}) + \nabla \cdot (\mu (\nabla \mathbf{u}^n + \nabla (\mathbf{u}^n)^T)) + \mathbf{f}^n + \frac{1}{\Delta t}\mathbf{u}^n\right)\Delta t - \frac{1}{\Delta t}\mathbf{u}^{n+1} \Delta t.
\end{equation}
Note that this is where this algorithm differs from Chorin's projection method.  Adding the additional $\frac{1}{\Delta t} \mathbf{u}$ terms to the equation allows the user to apply the small Reynolds Number and still obtain a useful equation.  The Reynolds Number is now taken to be zero, so the scheme applies to \ref{reducedNS}:

\begin{equation}
0 = P(\nabla \cdot (\mu (\nabla \mathbf{u}^n + \nabla (\mathbf{u}^n)^T)) + \mathbf{f}^n + \frac{1}{\Delta t}\mathbf{u}^n)\Delta t - \frac{1}{\Delta t}\mathbf{u}^{n+1} \Delta t.
\end{equation}
Solving for $u^{n+1}$ and distributing the $\Delta t$ into the $P$ operator yields

\begin{equation}
\label{step3Unsimplified}
\mathbf{u}^{n+1}  = P(\nabla \cdot (\mu (\nabla \mathbf{u}^n + \nabla (\mathbf{u}^n)^T))\Delta t + \Delta t  \mathbf{f}^n + \mathbf{u}^n).
\end{equation} 
Denote $\mathbf{u}^*$ as

\begin{equation}
\label{uStar}
\mathbf{u}^*  = \mathbf{u}^n + \Delta t(\nabla \cdot (\mu (\nabla \mathbf{u}^n + \nabla (\mathbf{u}^n)^T)) + \mathbf{f}^n) 
\end{equation} 
Then $\mathbf{u}^*$ can be substituted into equation \ref{step3Unsimplified} to recover

\begin{equation}
\label{uNextP}
\mathbf{u}^{n+1}  = P(\mathbf{u}^*).  
\end{equation}

Now the relationship between $\mathbf{u}^*$ and $p$ must be established.  Subtract equation \ref{projBothSides} from equation \ref{fullNSRearranged}:

\begin{equation}
\begin{aligned}
(\mathbf{I} - P)Re(\mathbf{u}^n_t) &= (\mathbf{I} - P)(-Re(\nabla \mathbf{u}^n \cdot \mathbf{u}^n) - \nabla p^n + \nabla \cdot (\mu (\nabla \mathbf{u}^n + \nabla (\mathbf{u}^n)^T)) + \mathbf{f}^n)\\
\implies Re\mathbf{u}^n_t - ReP(\mathbf{u}^n_t) &= (\mathbf{I} - P)(-Re(\nabla \mathbf{u}^n \cdot \mathbf{u}^n) - \nabla p^n + \nabla \cdot (\mu (\nabla \mathbf{u}^n + \nabla (\mathbf{u}^n)^T)) + \mathbf{f}^n)\\
\implies 0 &= (\mathbf{I} - P)(-Re(\nabla \mathbf{u}^n \cdot \mathbf{u}^n) - \nabla p^n + \nabla \cdot (\mu (\nabla \mathbf{u}^n + \nabla (\mathbf{u}^n)^T)) + \mathbf{f}^n)\\
\implies 0 &= -Re(\nabla \mathbf{u}^n \cdot \mathbf{u}^n) - \nabla p^n + \nabla \cdot (\mu (\nabla \mathbf{u}^n + \nabla (\mathbf{u}^n)^T)) + \mathbf{f}^n \\
&- P(-Re(\nabla \mathbf{u}^n \cdot \mathbf{u}^n) - \nabla p^n + \nabla \cdot (\mu (\nabla \mathbf{u}^n + \nabla (\mathbf{u}^n)^T)) + \mathbf{f}^n)\\
\implies \nabla p^n &= (I-P)(-Re(\nabla \mathbf{u}^n \cdot \mathbf{u}^n) + \nabla \cdot (\mu (\nabla \mathbf{u}^n + \nabla (\mathbf{u}^n)^T)) + \mathbf{f}^n)
\end{aligned}
\end{equation}
where $P(\mathbf{u}^n_t) = \mathbf{u}^n_t$ and $P(\nabla p) = 0$ are used.  Take $Re \rightarrow 0$ to obtain

\begin{equation}
\nabla p^n = (I-P)(\nabla \cdot (\mu (\nabla \mathbf{u}^n + \nabla (\mathbf{u}^n)^T)) + \mathbf{f}^n).
\end{equation}
Similar to before, $\frac{1}{\Delta t}u^n$ is added to both the P and identity operators.  The equation then becomes

\begin{equation}
\nabla p^n = (I-P)(\nabla \cdot (\mu (\nabla \mathbf{u}^n + \nabla (\mathbf{u}^n)^T)) + \mathbf{f}^n + \frac{1}{\Delta t}\mathbf{u}^n).
\end{equation}
Multiply by $\frac{\Delta t}{\Delta t}$:

\begin{equation}
\nabla p^n = \frac{1}{\Delta t}(I-P)(\Delta t(\nabla \cdot (\mu (\nabla \mathbf{u}^n + \nabla (\mathbf{u}^n)^T)) + \mathbf{f}^n) + \mathbf{u}^n)
\end{equation}
The interior is now replaced with $\mathbf{u}^*$:

\begin{equation}
\label{pIPuStar}
\nabla p^n = \frac{1}{\Delta t}(I-P)(\mathbf{u}^*).
\end{equation}
Next, apply equation (\ref{uNextP}):

\begin{equation}
\nabla p^n  = \frac{\mathbf{u}^* - \mathbf{u}^{n+1}}{\Delta t}
\end{equation}
Taking the divergence of both sides causes $\mathbf{u}^{n+1}$ to vanish due to the divergence-free condition:

\begin{equation}
\begin{aligned}
\Delta p^n  &= \frac{1}{\Delta t} \nabla \cdot (\mathbf{u}^* - \mathbf{u}^{n+1})\\
\implies \Delta p^n &= \frac{1}{\Delta t} \nabla \cdot \mathbf{u}^*
\end{aligned}
\end{equation}
%
%
%
The system is complete.  The Projection Method is now formally defined as the following three-step algorithm:

\begin{enumerate}
\item Obtain $\mathbf{u}^*$ via
\begin{equation}
\label{step1ProjectionMethod}
\mathbf{u}^*  = \mathbf{u}^n + \Delta t(\nabla \cdot (\mu (\nabla \mathbf{u}^n + \nabla (\mathbf{u}^n)^T)) + \mathbf{f}^n).
\end{equation} 
\item Solve for $p^n$.  Note that boundary conditions are required for $p$:
\begin{equation}
\label{step2ProjectionMethod}
\Delta p^n = \frac{1}{\Delta t} \nabla \cdot \mathbf{u}^*
\end{equation}
\item Obtain $\mathbf{u}^{n+1}$ via
\begin{equation}
\label{step3ProjectionMethod}
\mathbf{u}^{n+1}  = \mathbf{u}^* - \Delta t \nabla p^n.
\end{equation}
\end{enumerate}
If the viscosity does not vary spatially, then equation (\ref{step1ProjectionMethod}) becomes

\begin{equation}
\mathbf{u}^*  = \mathbf{u}^n + \Delta t(\mu \Delta \mathbf{u}^n + \mathbf{f}^n).
\end{equation} 


\subsection{Modeling a Vesicle Membrane}

The circular vesicle membrane in fluid applies a force to the fluid around it, causing a jump in pressure across the membrane.  This force is modeled by a modified Dirac delta function that is applied in a neighborhood of the membrane and nowhere else.  Further details on this approach, known as the Continuous Surface Force approach, can be studied in papers such as \cite{continuousSurfaceForce}.
\par
\begin{figure}
	\begin{center}
		\includegraphics[width=13cm]{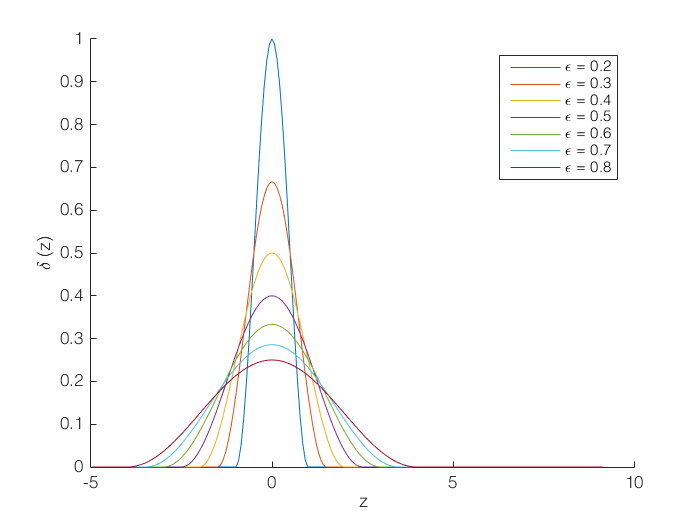}
	\end{center}
	\caption{Modified Dirac Delta Function}
	\label{diracGraph}
\end{figure}
\begin{figure}
	\begin{center}
		\includegraphics[width=13cm]{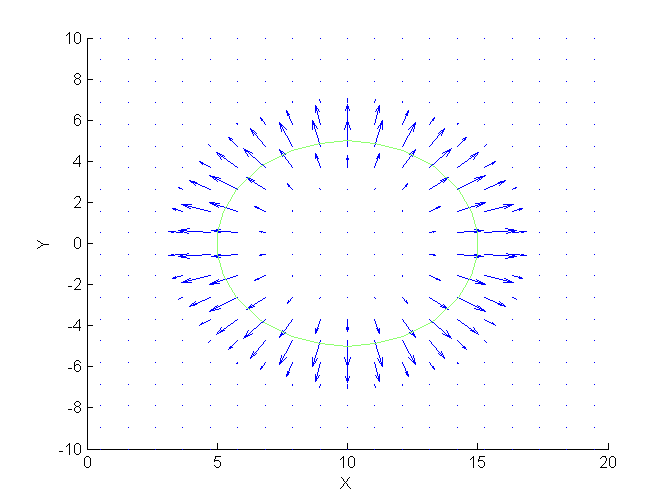}
	\end{center}
	\caption{Force from the Membrane}
	\label{forceGraph}
\end{figure}

The distance of the vesicle wall from the left edge of the domain is denoted as $L$, and the distance of the vesicle wall from its center is denoted as $R$.  Thus, the center of the vesicle is located at coordinates $(R+L, 0)$.  The signed distance from the vesicle membrane, denoted $z$, is then

\begin{equation}
\label{z}
z = \sqrt{(x-(R+L))^2 + y^2} - R.
\end{equation}
The modified Dirac Delta function is defined as

\begin{equation}
\label{dirac}
\delta(z) =
\begin{cases}
\frac{1 + \cos(\frac{\pi z}{\epsilon})}{2\epsilon} & \text{     if } -\epsilon \leq z \leq \epsilon \\
0 & \text{     otherwise}
\end{cases}.
\end{equation}
Graphically, this will create a smooth jump upward at the membrane with a height of $\frac{1}{\epsilon}$ and a width of $2\epsilon$, as shown in figure \ref{diracGraph} for varying values of $\epsilon$.
The actual value of $\epsilon$ in the algorithm is chosen to be $\frac{R}{2}$ to avoid numerical issues.  If this value is too small, the jump will be under-resolved and the discretized variables will skip right over the force.  If it is too big, then the force from the membrane will be spread out too far and will not properly capture the physics.  Since $z$ represents the distance from the membrane wall, $\nabla z$ points in the direction of maximum increase, which is always normal to the membrane.  Thus, multiplying $\delta(z)$ by $\nabla z$ will create a force vector that points outward from the membrane.  Noting that $1/R$ is the curvature ($\kappa$) of the membrane, Young-Laplace's Law gives that the force should be $\kappa \hat{n} \delta(z)$:

\begin{equation}
\mathbf{f} = \frac{1}{R} \delta(z) \nabla z
\end{equation}
The gradient of $z$ is calculated as follows:

\begin{equation}
\nabla z = 
\begin{bmatrix} 
\frac{x - (R + L)}{\sqrt{(x - (R + L))^2 + y^2 + \tilde{\epsilon}}} \\ 
\frac{y}{\sqrt{(x - (R + L))^2 + y^2 + \tilde{\epsilon}}} 
\end{bmatrix}
\end{equation}
where $\tilde{\epsilon}$ represents the small floating point value returned by the \verb|eps| Matlab command in order to prevent division by zero when the $(x,y)$ coordinate lies exactly on the membrane.  A graph of the force using $R = 5$, $L = 5$, $x \in (0,20)$, $y \in (-10,10)$ and $\epsilon = \frac{R}{2}$ is shown in Figure \ref{forceGraph}.  Dirichlet conditions are used in this scenario, setting $p$, $u$, and $v$ to $0$ at all of the boundaries.  A contour plot shows where $z=0$ to identify the membrane.

\subsection{Analytic Solution for the Vesicle Membrane Problem}

If a spatially-constant viscosity is assumed, the decoupled system of equations from \ref{decoupledSystem1} to \ref{decoupledSystem3} can be used.  Equations \ref{decoupledSystem2} and \ref{decoupledSystem3} can be simplified to provide the quickest route to the solution for $p$.  An ansatz is made such that $\mathbf{u}$ is identically zero, since the membrane is circular.  Under this assumption, $\Delta \mathbf{u}$ vanishes and equations \ref{decoupledSystem2} and \ref{decoupledSystem3} become

\begin{equation}
\label{pf}
\nabla p = \mathbf{f}.
\end{equation}
Because the fluid is stationary, the pressure is the only non-zero variable that needs to be solved.  Since the membrane force acts radially outward with the magnitude depending only on $z$, it can be assumed that $p(x,y) = p(r)$.  This implies converting to polar coordinates to obtain a solution.  Center the polar coordinates at the center of the vesicle, and define $r$ as the distance from that center.  With this definition, the signed distance from the membrane, $z$, becomes

\begin{equation}
z = r - R,
\end{equation}
since R is the distance from the center of the vesicle to the membrane.  The gradient of a function $f$ in polar, assuming that $\frac{\partial f}{\partial \theta} = 0$, is $\nabla f = f_r \hat{r}$.  Thus, the gradient of $z$ in this coordinate system is trivial: $\nabla z = \hat{r}$.  The modified Dirac Delta Function, as defined in equation \ref{dirac}, when converted to polar becomes

\begin{equation*}
\delta =
\begin{cases}
\frac{1}{2\epsilon}\left( 1 + \cos\left(\frac{\pi(r-R)}{\epsilon}\right)\right) & \text{     if } -\epsilon + R \leq r \leq R + \epsilon\\
0 &\text{     otherwise}
\end{cases}.
\end{equation*}
Thus, $\mathbf{f}$ within the region $-\epsilon + R \leq r \leq R + \epsilon$ becomes

\begin{equation}
\mathbf{f} = \frac{1}{2R\epsilon}\left( 1 + \cos\left(\frac{\pi(r-R)}{\epsilon}\right)\right) \hat{r}.
\end{equation}
Applying the polar gradient operator to $p$ yields $\nabla p = p_r \hat{r}$.  Using these identities, equation \ref{pf} becomes

\begin{equation}
 p_r \hat{r} = \frac{1}{2R\epsilon}\left( 1 + \cos\left(\frac{\pi(r-R)}{\epsilon}\right)\right) \hat{r}
\end{equation}
for $-\epsilon + R \leq r \leq R + \epsilon$.  Equating the vector components and integrating obtains

\begin{equation}
\begin{aligned}
\int p_r dr &= \int \frac{1}{2R\epsilon}\left( 1 + \cos\left(\frac{\pi(r-R)}{\epsilon}\right)\right) dr\\
\implies p &= \frac{1}{2R\epsilon}\left(r + \frac{\epsilon}{\pi}sin\left(\pi \frac{r-R}{\epsilon}\right)\right) + C_1
\end{aligned}
\end{equation}
for $-\epsilon + R \leq r \leq R + \epsilon$.  Note that 

\begin{equation}
\begin{aligned}
p_r &= 0\\
\implies p &= C_2
\end{aligned}
\end{equation}
for $|r - R| > \epsilon$.  If $r > R + \epsilon$, then $p = 0$ due to the specified boundary conditions.  The pressure must remain continuous at $p(R + \epsilon)$.  Thus,

\begin{equation}
\begin{aligned}
p(R + \epsilon) &= 0 = \frac{1}{2R\epsilon} (R + \epsilon + 0) + C_1\\
\implies C_1 &= \frac{-R - \epsilon}{2R\epsilon}\\
\implies p &= \frac{1}{2R}\left(\frac{r-R}{\epsilon} - 1 + \frac{1}{\pi}\sin\left(\pi\frac{r-R}{\epsilon}\right)\right).
\end{aligned}
\end{equation}
for $-\epsilon + R \leq r \leq R + \epsilon$.  If $r < R - \epsilon$, the pressure must remain continuous at $p(R - \epsilon)$.  Thus,

\begin{equation}
\begin{aligned}
p(R - \epsilon) &= C_2 = \frac{1}{2R}\left(\frac{R - \epsilon - R}{\epsilon} - 1 + 0\right)\\
\implies C_2 &= -\frac{1}{R}.
\end{aligned}
\end{equation}
The substitution $z = r - R$ is used to convert back to cartesian coordinates and produces $p$ within the region near the membrane:
\begin{equation}
p = \frac{-1}{2R}\left(1 - \frac{z}{\epsilon} - \frac{1}{\pi}\sin\left(\pi\frac{z}{\epsilon}\right)\right), \text{     if } -\epsilon \leq z \leq \epsilon.
\end{equation}
Thus, the analytic solution for the circular vesicle in fluid is

\begin{equation}
\mathbf{u} = 0
\end{equation}

\begin{equation}
p = 
\begin{cases}
\frac{-1}{2R}\left(1 - \frac{z}{\epsilon} - \frac{1}{\pi}sin\left(\pi\frac{z}{\epsilon}\right)\right)& \text{     if } -\epsilon \leq z \leq \epsilon\\
-\frac{1}{R}& \text{     if } z < -\epsilon\\
0 & \text{     if } z > \epsilon
\end{cases}.
\end{equation}


\subsection{Staggered Grid for Projection Method}

\begin{figure}
	\begin{center}
		\includegraphics[width=10cm]{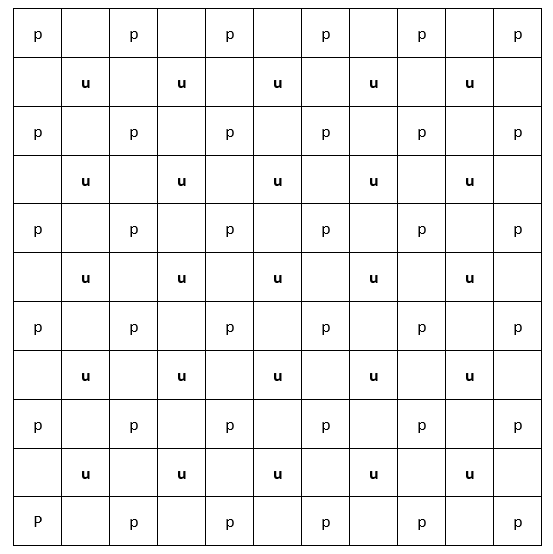}
	\end{center}
	\caption{Descretized Grid for the Projection Method}
	\label{projStaggered}
\end{figure}

A different staggered grid is used for this algorithm than what was used for the Saddle-Point method, but the motivation is the same:  to prevent a checkerboard pattern caused by skipping information from adjacent cells.  The $\mathbf{u}$ and $p$ cells are arranged as shown in figure \ref{projStaggered}.  The distance $\Delta x$ is defined as the distance between neighboring cells of the same type.  
\par
Discretization is done at each cell, so cells using values of cells of a different type must use the average values of those cells around them.  Neighboring cells of a different type are a distance of $\frac{\Delta x}{2}$ away in the horizontal direction and $\frac{\Delta y}{2}$ away in the vertical direction.  To derive a finite-difference approximation to $p_x$ at a $\mathbf{u}$ node, Taylor expand $p(x \pm \frac{\Delta x}{2}, y)$ and $p(x, y \pm \frac{\Delta y}{2})$ about $(x,y)$:

\begin{equation}
\begin{aligned}
p(x + \frac{\Delta x}{2},y) = p(x,y) + \frac{\Delta x}{2} \frac{\partial p(x,y)}{\partial x} + O(\Delta x^2)\\
p(x - \frac{\Delta x}{2},y) = p(x,y) - \frac{\Delta x}{2} \frac{\partial p(x,y)}{\partial x} + O(\Delta x^2)\\
\end{aligned}
\end{equation}
Subtracting these two equations and simplifying yields:

\begin{equation}
p_x = \frac{[p(x + \frac{\Delta x}{2},y) - p(x - \frac{\Delta x}{2},y)]}{ \Delta x} + O(\Delta x^2).
\end{equation}
Thus, the first derivative may be calculated using neighboring cells with a total distance of $\Delta x$ between them.  But these cells directly to the right and left of a $\mathbf{u}$ cell do not actually contain $p$ values on the specified grid.  Averaging the values of these derivatives above and below the actual $\mathbf{u}$ cell is therefore used to yield the true formula for $p_x$ at a $\mathbf{u}$ cell:

\begin{equation}
p_x \approx \frac{1}{2}\left(\frac{p(x + \frac{\Delta x}{2},y + \frac{\Delta y}{2}) - p(x - \frac{\Delta x}{2},y + \frac{\Delta y}{2})}{ \Delta x} + \frac{p(x + \frac{\Delta x}{2},y - \frac{\Delta y}{2}) - p(x - \frac{\Delta x}{2},y - \frac{\Delta y}{2})}{ \Delta x}\right)
\end{equation}
Using subindices, the finite-difference formula is

\begin{equation}
p_x \approx \frac{1}{2}\left(\frac{p_{i+1,j+1} - p_{i,j+1}}{ \Delta x} + \frac{p_{i+1,j} - p_{i,j}}{ \Delta x}\right).
\end{equation}
The same logic is applied for $p_y$:

\begin{equation}
p_x \approx \frac{1}{2}\left(\frac{p_{i+1,j+1} - p_{i+1,j}}{ \Delta y} + \frac{p_{i,j+1} - p_{i,j}}{ \Delta y}\right)
\end{equation}
At a $p$ cell, the $\mathbf{u}$ derivatives are also calculated using the same approach, with a difference of one in the indexing due to the orientation of the grid cells:

\begin{equation}
u_x \approx \frac{1}{2}\left(\frac{u_{i,j} - u_{i-1,j}}{ \Delta x} + \frac{u_{i,j-1} - u_{i-1,j-1}}{ \Delta x}\right)
\end{equation}

\begin{equation}
v_y \approx \frac{1}{2}\left(\frac{v_{i,j} - v_{i,j-1}}{ \Delta x} + \frac{v_{i-1,j} - v_{i-1,j-1}}{ \Delta x}\right)
\end{equation}
Other operators that do not involve cells of a different type are unchanged, using the normal finite-difference formulas. 

\subsection{Convergence Testing}

\begin{figure}
	\begin{center}
		\includegraphics[width=10cm]{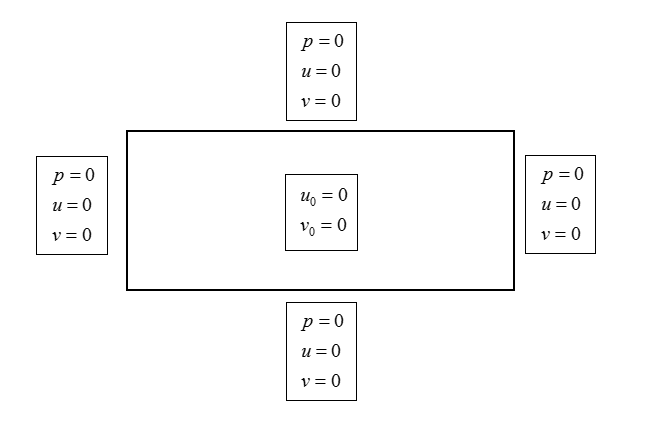}
	\end{center}
	\caption{Initial and Boundary Conditions Used for the Projection Method on the Vesicle Force Problem}
	\label{projectionBC}
\end{figure}

Since the Projection Method requires $\mathbf{u}^n$ to compute $\mathbf{u}^{n+1}$, an initial $\mathbf{u}^0$ is needed.  For the circular vesicle in fluid problem, the initial condition on $\mathbf{u}$ will be that it is just zero everywhere.  Homogeneous Dirichlet boundary conditions are imposed so that $\oint\limits_{\partial \Omega} p \mathbf{u}_t \cdot \hat{n} \text{ } dS = 0$, ensuring that the inner product between $\mathbf{u}_t$ and $\nabla p$ is zero.  The pressure is expected to approach zero as the distance from the vesicle approaches infinity.  Thus, the pressure will be zero at the boundaries as well.  These boundary and initial conditions are illustrated in Figure \ref{projectionBC}.

The numerical solution obtained using the Projection Method produces the steady-state solution after one time step of $u = 0$, $v = 0$, and pressure with a rapid gradient near the wall of the vesicle and a region of negative pressure within the vesicle, as expected from the analytical solution.  The parameter values chosen are height $=20$, width $=20$, $M = 50$ cells in each spatial direction, $R = 5$, $L = 5$, and $\mu = 1$.  The discrete L2 error is plotted versus $\Delta x$ on a log-log scale as before.  The L2 error shows second-order convergence, as expected from the integration approximation used in (\ref{integralApprox}).  
\par
During the derivation, when $\frac{1}{\Delta t} \mathbf{u}$ is added to the inside and outside of the projection operator (\ref{addInside}), the constant $\frac{1}{\Delta t}$ is necessary in order to stabilize the numerical method.  This was not originally part of the derivation, but was added when it was realized that the method did not converge.  The value $\Delta t = \Delta x^2$ is used in the algorithm to ensure stability.
\par
\begin{figure}
	\begin{center}
		\includegraphics[width=13cm]{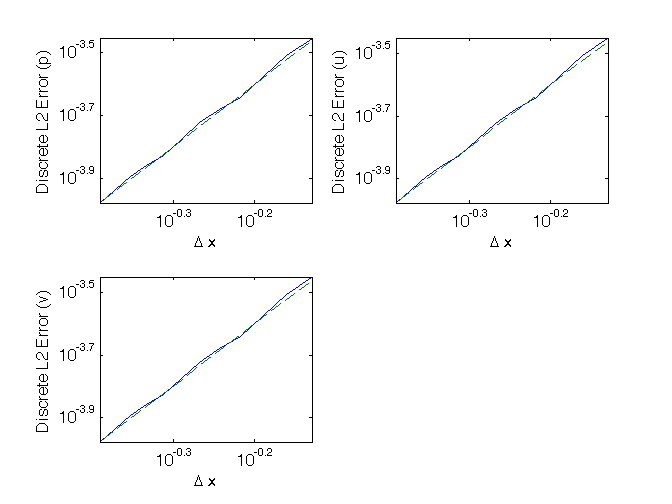}
	\end{center}
	\caption{The graphs of the discrete L2 error for u, v, and p, respectively, on a log-log scale using the Projection Method.  The dashed line is of slope 2 on the log-log scale as a reference line to show second order convergence.}
	\label{projectionL2E}
\end{figure}
\begin{figure}
	\begin{center}
		\includegraphics[width=13cm]{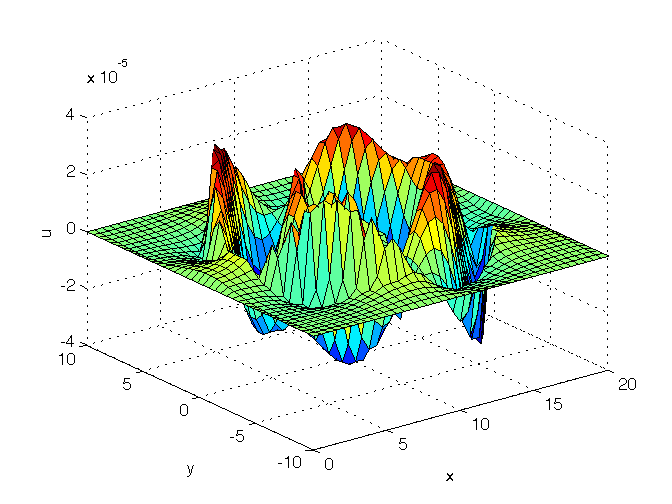}
	\end{center}
	\caption{The horizontal velocity solution obtained from the Projection Method.  It should be zero everywhere, and the magnitude does converge to zero as $O(\Delta x^2)$}
	\label{projectionU}
\end{figure}
\begin{figure}
	\begin{center}
		\includegraphics[width=13cm]{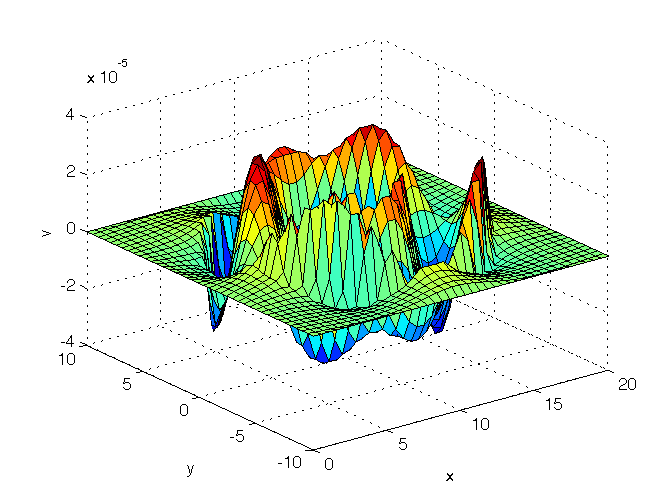}
	\end{center}
	\caption{The vertical velocity solution obtained from the Projection Method.  It should be zero everywhere, and the magnitude does converge to zero as $O(\Delta x^2)$}
	\label{projectionV}
\end{figure}
\begin{figure}
	\begin{center}
		\includegraphics[width=13cm]{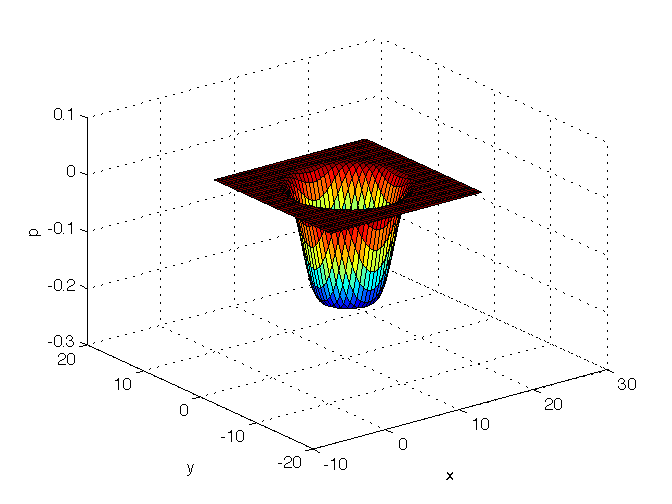}
	\end{center}
	\caption{The pressure solution obtained from the Projection Method.  It shows zero pressure outside of the membrane, a sharp gradient near the vesicle wall, and a region of negative pressure inside the cell.}
	\label{projectionP}
\end{figure}

The computational results are shown in Figures \ref{projectionL2E} through \ref{projectionP}.  Similar convergence results are produced when using the Decoupling or Saddle-Point methods on the same parameters.

\section{Comparisons}

\subsection{Time Analysis}
\label{TimeAnlysisSection}

In each of the Saddle-Point, Decoupling, and Projection methods, the linear system $\mathbf{A} \mathbf{x} = \mathbf{b}$ is solved using the Matlab \verb|"\"| operator.  The number of these solves and the size of $\mathbf{A}$ varies between the different algorithms, impacting their performance times.
\par
In a worst-case scenario, \verb|"\"| is considered as equivalent to Gaussian elimination in order to analyze the number of floating-point operations (FLOPS) required to solve the equation.  In converting $\mathbf{A}$ to Reduced Row Echelon form, each row must be multiplied by a specific coefficient to eliminate the leading coefficients of each row below it.  This will result in $(n-1)^2 + (n-2)^2 + ... + 4 + 1$ multiplications, where $n$ is the number of rows.  Applying the sum of squares formula, this result becomes $\frac{n(n-1)(2n-1)}{2}$ multiplications.  Each of these multiplied rows must be added to the other rows to elminate the leading coefficients, resulting in $\frac{n(n-1)(2n-1)}{2}$ additions.  These two major operations together result in the total number of floating point operations to solve the equation being $O(n^3)$.
\par
The Decoupling method involves three different solves of $\mathbf{A} \mathbf{x} = \mathbf{b}$:  One each for $u$, $v$, and $p$.  Each of them inverts a matrix of size $M^2 \times M^2$, since $\mathbf{x}$ represents the vectorized two-dimensional vector on an $M \times M$ grid for the particular variable being solved.  Thus, the execution time should scale like $3 \cdot O(M^6)$, since there are three solves.
\par
The Saddle-Point method involves one solve of $\mathbf{A} \mathbf{x} = \mathbf{b}$, where $\mathbf{A}$ is of size $3M^2 \times 3M^2$, because all of the discretized values of $u$, $v$, and $p$ are stacked into one single vector.  In this case, $n$ is three times larger than it is in the Decoupling Method.  Since the solve is $O(n^3)$, this will result in a staggering $27 \cdot O(M^6)$ for its execution time.
\par
The Projection method performs three steps on its path to solving $p$, $u$, and $v$ for one time step, as shown in equations \ref{step1ProjectionMethod} through \ref{step3ProjectionMethod}.
Steps one and three of the Projection Method do not involve any matrix inversions, so there is a significant time savings in this algorithm.  An assignment must be done for each of the $n=M^2$ discretized points, resulting in $2 \cdot O(M^2)$ operations to complete those two steps.  Step two involves a matrix inversion to solve for $p$.  Since the matrix only needs to solve for $p$, it will be of size $M^2 \times M^2$.  Thus, the overall complexity of this algorithm is just $O(M^6)$.  This should result in a performance increase over the Decoupling Method, which is $3 \cdot O(M^6)$, and a substantial increase over the Saddle-Point Method, which is $27 \cdot O(M^6)$.

\subsection{Execution Times}

\begin{figure}
	\begin{center}
		\includegraphics[width=13cm]{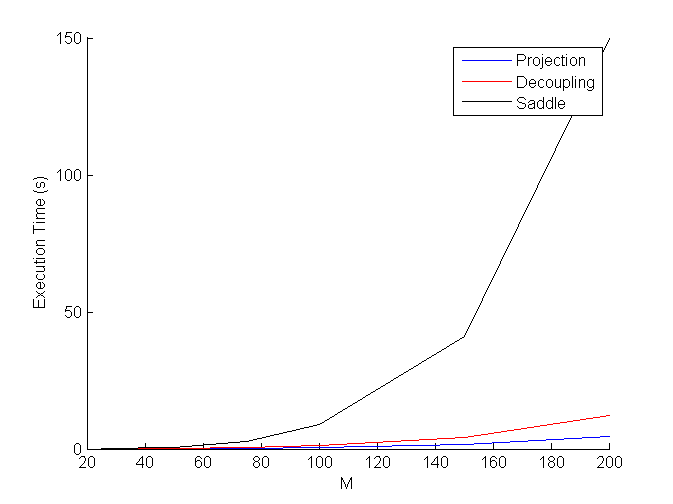}
	\end{center}
	\caption{Graph of execution time vs. the number of discretization points for Decoupling, Saddle-Point, and Projection methods.}
	\label{allTime}
\end{figure}

\begin{figure}
	\begin{center}
		\includegraphics[width=13cm]{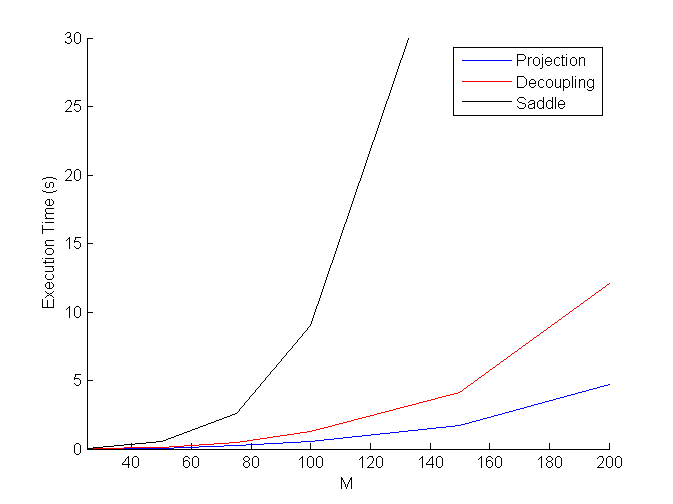}
	\end{center}
	\caption{Graph of execution time vs. the number of discretization points for Decoupling, Saddle-Point, and Projection methods, zoomed in.}
	\label{allTime2}
\end{figure}

\begin{figure}
	\begin{center}
		\includegraphics[width=13cm]{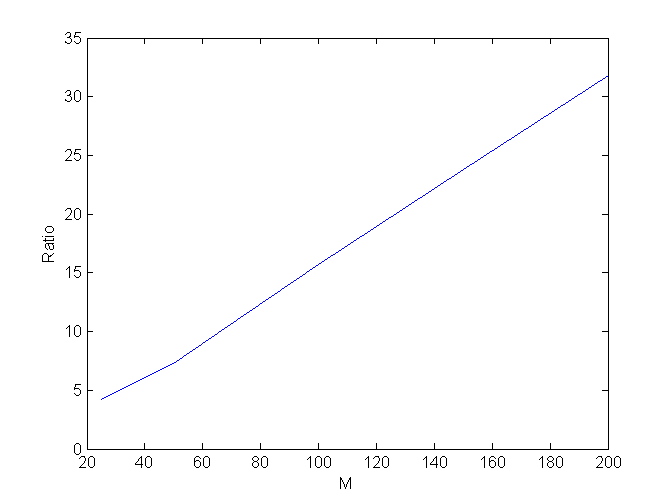}
	\end{center}
	\caption{Graph of the ratios of the execution times for various discretization values between the Saddle-Point method and the Projection Method:  Saddle-Point method time divided by Projection method time.}
	\label{PSRatios}
\end{figure}

\begin{figure}
	\begin{center}
		\includegraphics[width=13cm]{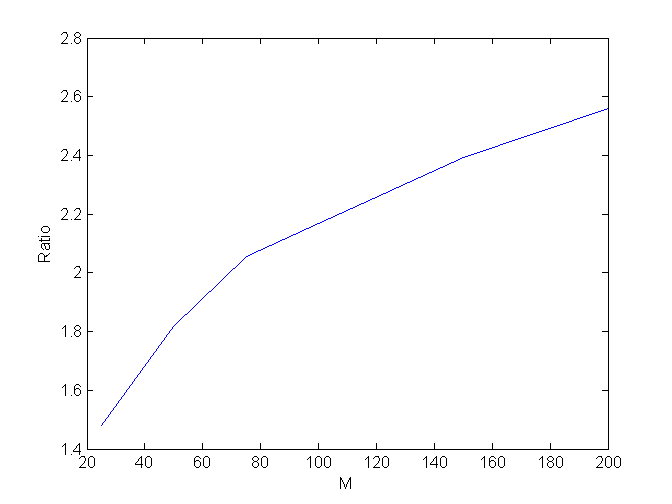}
	\end{center}
	\caption{Graph of the ratios of the execution times for various discretization values between the Decoupling Method and the Projection Method:  Decoupling method time divided by Projection method time.}
	\label{PDRatios}
\end{figure}

Execution times are obtained in Matlab and measured as an average of 10 runs to mitigate variation in runtime due to background processes.  Each measurement is taken with varying discretization values:  $M=25$, $M=50$, $M=75$, $M=100$, $M=150$, and $M=200$.  Since the Projection Method iterates forward in time, one time step is solved to create an accurate comparison, although it is capable of solving as many time steps as necessary and remains stable to second order throughout each time step.  The other parameters used are the same as during the convergence testing:  height $=20$, width $=20$, $R = 5$, $L = 5$, and $\mu = 1$.  The results are shown in Figures \ref{allTime} through \ref{PDRatios}, which support the hypothesis in section \ref{TimeAnlysisSection} that the Projection Method will be faster than the Decoupling or Saddle-Point methods.


\begin{table*}
	\centering
		\begin{tabular}{c | c | c | c}
			\hline
			Discretization Points (M) & Decoupling (s) & Saddle-Point (s) & Projection (s)\\ \hline
			25 &  0.0184 &  0.0519 &  0.0124\\ \hline
			50 &  0.1301 &  0.5195 &  0.0715\\ \hline
			75 &  0.4659 &  2.6204 &  0.2266\\ \hline
			100 &  1.2452 & 9.0459 & 0.5742\\ \hline
			150 &  4.1129 & 40.9595 & 1.7194\\ \hline
			200 & 12.0575 & 149.6981 & 4.7079\\ \hline
		\end{tabular}
	\caption{Execution times in seconds for the Decoupling, Saddle-Point, and Projection methods.}
	\label{tab:RunTimes}
\end{table*}

\section{Conclusions}


The existing methods analyzed for solving Stokes Flow are the Saddle-Point method and the Decoupling method.  The Decoupling method is much faster in execution time than the Saddle-Point method, but can only be used when the viscosity is spatially constant.  For problems like the flow of red blood cells under varying solute concentrations, spatially constant viscosity is likely not valid.  The Chorin projection method is modified for use in solving the Stokes Flow with spatially varying viscosity.
\par
As demonstrated in the Time Analysis section, the Projection Method should theoretically run three times faster than the Decoupling method and $27$ times faster than the Saddle-Point method.  Upon observing figure \ref{PSRatios}, the ratio of execution time between the Projection and Saddle-Point method seems to be dependent upon the number of discretization points $M$, taking on the ratio of $27$ around $M = 180$ but then exceeding it as $M$ grows.  Figure \ref{PDRatios} shows the ratio asymptotically approaching $3$, which matches the theoretical results.  Possible confounding factors that cause these results to not match perfectly with the theoretical time analysis include overhead from other operations in the code or operating system and the Matlab \verb|"\"| operator having a best case $O(n)$ and worst case $O(n^3)$ complexity. 
\par
These results are very promising.  The Projection Method possesses the ability to solve the Stokes Flow with varying viscosity, but is also significantly faster than both of the existing methods.

\section{Github Repositories}

The following repositories contain the Matlab code mentioned in this project:
\begin{enumerate}
\item Projection Method on Vesicle Membrane:
\par
https://github.com/rhermle/ProjectionMethod
\item Saddle-Point Method on Vesicle Membrane:
\par
https://github.com/rhermle/SaddleVesicle
\item Decoupling Method on Vesicle Membrane:
\par
https://github.com/rhermle/DecouplingVesicle
\item Saddle-Point Method with Checkerboard Error:
\par
https://github.com/rhermle/SaddleCheckerboard
\item Saddle-Point Method on Fluid in a Pipe:
\par 
https://github.com/rhermle/SaddlePipe
\item Decoupling Method on Fluid in a Pipe:
\par
https://github.com/rhermle/DecouplingPipe

\end{enumerate}

\end{document}